 \documentclass[preprint2]{aastex62}

\shorttitle{Eta Car UV} 
\shortauthors{Davidson et al.}

\begin{document}

\title{ ETA CARINAE'S DECLINING OUTFLOW SEEN IN THE UV, 
   2002-2015\footnote{ 
       Based on observations made with the NASA/ESA Hubble Space 
       Telescope, which is operated by the Association of 
       Universities for Research in Astronomy, Inc., under 
       NASA contract NAS 5-26555.} }

  \author{Kris Davidson}
  \affiliation{Minnesota Institute for Astrophysics, 116 Church St SE, 
     University of Minnesota, Minneapolis, MN 55455} 
  \author{Kazunori Ishibashi}  
  \affiliation{Graduate School of Science, Nagoya University, 
     Nagoya, 464-8602, Japan} 
  \author{John C. Martin} 
  \affiliation{Barber Observatory, University of Illinois, 
      Springfield, IL, 62703}
  \author{Roberta M. Humphreys} 
  \affiliation{Minnesota Institute for Astrophysics, 116 Church St SE, 
     University of Minnesota, Minneapolis, MN 55455}

\begin{abstract}
Existing HST UV data offer many previously neglected clues 
to $\eta$ Car's behavior since 2000.  Here we examine a subset     
of observations with diverse results.   
(1) The star's rapid change of state is confirmed by major changes 
in UV absorption lines,  circumstellar extinction, and other features.  
(2) \ion{N}{3}] $\lambda$1750 is one of the two most luminous emission 
features  in  $\eta$ Car's observable spectrum, comparable to H$\alpha$.
This and other 
semi-forbidden lines are useful because they have no P Cyg absorption. 
(3) \ion{N}{3}] multiplet ratios provide the first direct diagnostic 
of gas densities in $\eta$ Car's outflow.  They strongly suggest 
that high-excitation lines originate in condensations within 
the colliding-wind shocked region.  The parameters imply that 
published models have not adequately represented the essential 
small size scales.     
(4) In 2002-2004, a very large amount of \ion{N}{3}] emission 
had anomalous Doppler velocities from $+400$ to $+1200$ km s$^{-1}$.     
This is a mystery;  we conjecture that it may have resulted from a 
burst of mass ejection in the 2003.5 periastron event. 
Various other effects are also difficult to explain and merit further 
investigation. 
\end{abstract}    

    \vspace{1mm}

\section{Introduction}  
\label{sec:intro}

Eta Carinae's unique role in stellar astrophysics arises from 
three circumstances: 
\begin{enumerate} 
  \item Its primary star is the only supernova impostor  
  or giant-eruption survivor that can be studied in detail.  
  \item Its dense wind has changed dramatically in the past 20 years.    
  \item Extraordinary effects occur at 5.5-year intervals during 
  periastron passages of a companion star.   
\end{enumerate}  
Item 1 is fundamental because the basic instability remains 
mysterious after many years of study; while 2 and 3 can provide 
clues to the star's internal recovery following its Great Eruption  
observed in 1830--1860.  For general information, see reviews by  
many authors in \citet{dh12}.  

    \vspace{1mm}  

The Hubble Space Telescope (HST) revolutionized this topic, because all 
ground-based photometry and spectroscopy of $\eta$ Car is severely 
contaminated by ejecta at $r \sim$ 300-1500 AU, seen less than 
0.5 arcsec from the star. 
(See Fig.\/ 1 in \citealt{da15}, Fig.\/ 5 in \citealt{da95}, 
and discussions in \citealt{me11,me12}.) 
The Space Telescope Imaging Spectrograph (STIS) has played the 
largest role in this story, using wavelengths longer than 250 nm.  
In this paper we explore FUV wavelengths between 160 and 230 nm -- 
a spectral region that is physically different as noted in   
{\S}\ref{sec:contin} below.  

    \vspace{1mm} 

For stellar astrophysics the {\it major\/} problem 
of $\eta$ Car is the nature of its instability and its Great 
Eruption (see \citealt{da12} for a semi-theoretical account).   
In that sense, the most crucial development in this topic 
since 2000 has been the rapid trend of changes in the star.  
Recent publications tend to focus instead on geometries  
of the colliding winds, the binary orbit, and other matters  
that are less fundamental (e.g., \citealt{gu16,ri16,te16,we16} 
and many others).  That is true of {\S}\ref{sec:niii1750} and 
{\S}\ref{sec:emlines} in this paper as well, 
but our motivation differs.    Rather than viewing the complicated 
gas flows as the main topic in their own right, we hope that their 
secular changes  will eventually lead to information about the 
disequilibrium structure of the post-eruption star.  Observed 
long-term spectroscopic  
{\it trends\/} are therefore critical.  The goal of linking 
them to the star's internal structure has not been approached 
yet, but no other known object can offer comparable views into   
a supernova impostor. 

    \vspace{1mm} 

Here we explore a particular subset of UV observations, selected by a 
criterion noted in {\S}\ref{sec:data}.  They contain a  wide variety 
of results.  Some of these relate to $\eta$ Car's secular trend,   
while others involve the morphology and physics of gas flows in the 
system.  Several phenomena were not recognized earlier, and appear 
difficult to explain.    

    \vspace{1mm} 

Section 2 below is a brief description of the data set, and Section 3  
is an overview of the rapidly  growing UV brightness levels.   
In Section 4 we focus on the extraordinarily bright 
\ion{N}{3}] $\lambda$1750 multiplet which has several 
important implications.   In Section 5 we note the 
potential importance of other semi-forbidden lines, and Section 6  
contains an account of a few especially suggestive absorption  
features.  Finally, Section 7 is an attempt to summarize the 
resulting situation.


\section{STIS/MAMA/UV Observations and Data Reduction} 
\label{sec:data} 

    \vspace{1mm}

HST/STIS has played a unique role in our knowledge of $\eta$ Car for   
three reasons: (1) As noted above, it can separate the central star 
from the surprisingly bright ejecta located 0.1--0.5 arcsec away.  
(2) It allows UV observations.  
(3) It also provides unrivaled data homogeneity  
since 1998, immune to atmospheric effects and far more reproducible  
in spatial coverage than ground-based data. 

   \vspace{1mm}  
 \begin{figure}  
 \plotone{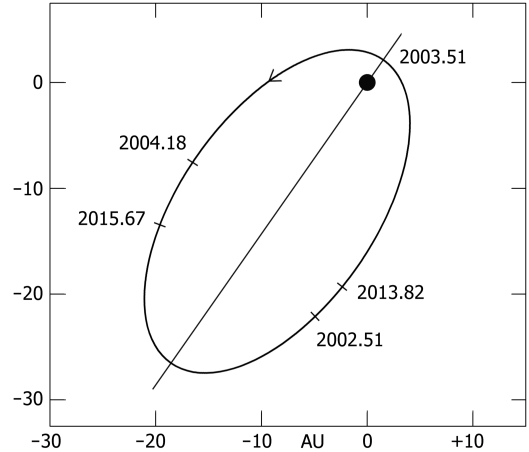}  
 \caption{ The 5.54-year orbit of $\eta$ Car's companion star, with dates  
    of UV observations discussed in this paper.  In most models our line 
    of sight passes upward in the figure, inclined 45$^\circ$ from the 
    orbital plane, but some authors advocate the opposite direction. 
    Here we assume eccentricity 0.85,  see refs.\ cited in the text. } 
 \label{fig:orbit} 
 \end{figure}

Here we are interested in secular trends at $\lambda < 2400$ {\AA},  
observable with the STIS/MAMA detectors.    
In order to minimize periodic variations related to the 
5.54-year orbit,  we chose pairs of observations that 
occurred  at similar orbital phases.   There are only 
two such pairs, shown in Figure \ref{fig:orbit}.  One of them includes 
$t =$ 2002.51 and 2013.82, about a year before periastron; while the other 
pair occurred at 2004.18 and 2015.67, after periastron.  Eta Car was 
not observed with STIS/MAMA in the intermediate years 2005--2012.     
Table \ref{tab:obslist} lists the four selected observations, plus another  
that occurred during the 2003.5 spectroscopic event near periastron.

   \vspace{1mm} 

\begin{deluxetable}{ccccc}   
    \tablenum{1}
\tablecaption{Times of HST/MAMA echelle observations used in this paper} 
\label{tab:obslist}  
\tablehead{ 
	\colhead{Year}  & \colhead{Calendar} & \colhead{MJD}  
	& \colhead{Phase\tablenotemark{a} }   
	& \colhead{Program\tablenotemark{b} }   }  
\startdata
J2002.51  &  2002-07-05  &  52459.9  &  0.8136  &  \ 9337  \\ 
J2003.51  &  2003-07-05  &  52825.0  &  0.9941  &  \ 9973  \\   
J2004.18  &  2004-03-06  &  53070.3  &  0.1153  &  \ 9973  \\  
J2013.82  &  2013-10-27  &  56591.8  &  0.8561  &   13377  \\   
J2015.67  &  2015-09-02  &  57267.3  &  0.1900  &   13789  \\ 
\enddata
\tablenotetext{a}{Phase in the 5.54-year orbit, with nominal period  
   2023.0 days and zeropoint at J1998.000.  Periastron is 
   ill-determined but most likely occurs in the phase range 
	0.988--0.998. }  
\tablenotetext{b}{HST/GO program number. P.I.'s were  
   K.\/ Davidson in 2002-2004 and A.\/ Mehner in 2013-2015. } 
\end{deluxetable}  

    \vspace{1mm}  

The STIS/MAMA/FUV detector sampled wavelength range 
1300--1700 {\AA} with echelle grating E140M, and the NUV 
detector sampled 1610--2360 {\AA} with grating E230M.  
Spectral resolution was equivalent to ${\Delta}v \sim 10$ km s$^{-1}$, 
much narrower than most of the spectral features.  In 2002-2004 
the integration times were around 2000 s with E140M, and 3500 s 
with E230M.  Eleven years later $\eta$ Car's brightening allowed  
shorter integrations, about 1000 s and 800 s respectively.  
Since the resulting data had many thousands of counts per {\AA} 
at most wavelengths,  statistical noise was much smaller than 
other sources of uncertainty that will be obvious below.     
The spectrograph aperture size was $0.2 \times 0.2$ arcsec for 
E140M, and $0.2 \times 0.3$ arcsec for E230M.  These were near the 
maximum size that can exclude emission from nearby ejecta. 
On the other hand,  \citet{hi06} noted that we must include 
as much as possible of a UV scattering halo whose effective 
diameter  exceeded 0.1 arcsec in 2002-2004.  Therefore 
the chosen aperture size was a reasonable compromise. 

     \vspace{1mm} 

Our data reduction methods essentially followed the standard STScI 
CALSTIS pipeline processing.  Spectral extractions presented in this 
paper were performed with the full aperture width and standard   
pipeline aperture corrections.  Strictly speaking the aperture 
corrections may be unreliable for $\eta$ Car due to the noticeable 
extent of its emission;  but there is not enough information 
to do much better.  
Judging from the near-absence of very narrow emission lines, and from 
spatial structure within the aperture width, the nearby slow-moving 
ejecta did not appreciably contaminate these data. 

    \vspace{1mm}  

Orbital positions at the times of these observations are sketched  
in Figure \ref{fig:orbit}.  According to most authors we view the 
orbit from below the bottom edge of the figure and roughly 45$^\circ$
out of its plane, but a nearly opposite line of sight has also been 
proposed, see refs.\ in \citet{ka16}.  As noted in 
{\S}\ref{sec:niii1750}  below, high-excitation species such as 
He$^+$ and N$^{++}$ should occur in and near a shocked zone where the 
two winds collide, between the two stars.  
Concerning the orbit parameters, see \citet{da01,da17}, 
{\S}6.2 in \citet{me11}, and    references therein.
None of our conclusions depend on their precise values.     

    \vspace{1mm} 

Regarding ``phase'' in the 5.5-year orbital cycle, we use the definition 
adopted 15 years ago for the $\eta$ Car Treasury Project:  
period = 2023.0 days exactly, and phase is zero (or alternatively an integer) 
at MJD 56883.0 = 2014 August 14, MJD 54860.0 = 2009 January 29, etc.  
This system has been used in most accounts of HST data on the central 
star, and reasons for it are outlined in an Appendix in \citet{me11}.     

   \vspace{1mm} 

Quoted wavelengths are in vacuum, and Doppler velocities are heliocentric.  
Eta Car's heliocentric systemic radial velocity is probably in the range 
0 to $-20$ km s$^{-1}$ \citep{da97}.  

   \vspace{4mm}  

\section{The UV Jungle, and Diminishing Extinction}   
\label{sec:contin} 
  
  \vspace{1mm}   

\begin{figure}   
 \plotone{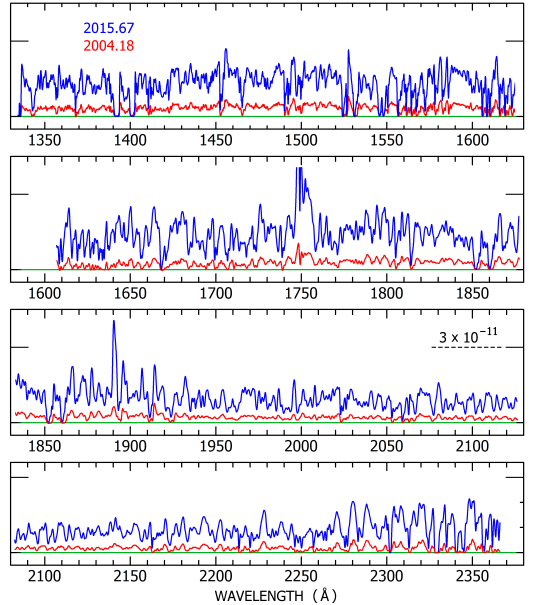}  
 \caption{UV spectrum observed at 2015.67, and also at 2004.18 to show 
    the apparent brightening.  Horizonal marks indicate flux level   
    $f_\lambda = 3 \times 10^{-11}$ erg cm$^{-2}$ s$^{-1}$ {\AA}$^{-1}$, 
    not corrected for extinction.  The spectral features visible here 
    are nearly all real, since the statistical noise level is too small 
    to discern at this scale. } 
 \label{fig:uvtracing} 
 \end{figure}

Figure \ref{fig:uvtracing} shows $\eta$ Car's UV spectrum at $t = 2015.67$. 
It is far  more complex than the spectral region $\lambda > 3500$ {\AA},    
cf.\ tracings in \citet{da95,da99a,hu99,hi01,hm12,na12}.  The UV continuum 
level cannot be clearly identified.  \citet{eb97}, \citet{wa99}, and 
\citet{hi01} compared $\eta$ Car's FUV spectrum to various known 
single-star winds, but later it was recognized that the hot companion 
star photoionizes some of the observed gas -- see {\S}6 in  
\citet{hu08}, and {\S}\ref{sec:niii1750} below.  
The secondary star itself is relatively faint at wavelengths 
examined here  \citep{me10a}.   

      \vspace{1mm} 

One might hope that the hundreds of features in $\eta$ Car's UV spectrum
can be  analyzed via numerical simulations 
\citep{hi01,gr12a,mad13,cl15}.  
However,  there are too many asymmetric emitting regions and 
small-scale processes, with far too many free parameters, for a reliable 
model with any set of computer codes available now or in the near 
future (see {\S}\ref{subsec:niiic} and {\S}\ref{subsec:sum2} below).  
We therefore concentrate on spectral features that depend chiefly 
on just a few physical processes. 

       \vspace{1mm}  


\subsection{A minor paradox}   
\label{subsec:contin_a}  

First, though, note that most features in this wavelength range 
do not depend on the same parameters as the violet-to-red spectrum.  
\citet{me10b} found that $\eta$ Car's \ion{Fe}{2} emission lines at 
$\lambda > 4000$ {\AA} weakened by factors of 2--5 between the years 
2000 and 2010, relative to the continuum.  In our UV data, however, 
the \ion{Fe}{2} forest appeared roughly the same in 2015 as it did 
in 2002.  This difference is not a contradiction, for the following 
reasons.  

  \vspace{1mm}    

\ion{Fe}{2} has about three dozen levels with $E_i < 3$ eV, mostly 
well populated because they are metastable.  They 
produce a few hundred UV absorption lines with 
substantial optical depths in $\eta$ Car's {\it outer} wind, at 
$r \sim$ 50--200 AU.  A typical absorption event is followed by 
re-emission either in the same line, or in another UV line with 
$\lambda < 3000$ {\AA}.   (Longer-wavelength transitions  
have much smaller probabilities.)  Many scattering events -- really 
absorption and re-emission -- may occur before the photon escapes.  
Consider first a simplified  case with only 
one spectral line.   In a stellar wind, it produces a P Cyg 
profile of the pure-scattering type:  some energy is removed from 
a narrow wavelength interval and transferred to slightly longer 
wavelengths.  If the 
line-center optical depth ${\tau}_0$ is larger than about  1.5, 
then the amplitude and width of the resulting line 
profile do not strongly depend on the actual value of ${\tau}_0$.  
In other words, if more than (say) four scattering events occur 
before photon escape, then the specific number of them has only 
a weak influence on the line's apparent strength.  Thus {\it a major 
change in the gas density does not greatly alter the appearance of 
the spectral feature.}  The line's quantitative details are affected, 
but not its basic nature.  

    \vspace{1mm}  
   
In reality each \ion{Fe}{2} absorption event may be followed  
by re-emission in a different UV transition,  but this fact merely 
re-distributes the scattering among a limited set of spectral lines, 
and the pure-scattering P Cyg concept remains valid for their average. 
In summary, we should not expect a conspicuous change in the general  
appearance of $\eta$ Car's ultraviolet \ion{Fe}{2} forest;  that would 
happen only if nearly all of the line optical depths ${\tau}_0$ fall 
below 1.5.     The same remarks apply to \ion{Ni}{2} and other complex 
species.

    \vspace{1mm} 

Sometimes the scattering sequence ends with a relatively unlikely 
non-UV decay, producing a photon with $\lambda > 4000$ {\AA} which 
escapes more easily.  This has a specific probability of occurring 
after each UV absorption event,   so the resulting amount of 
longer-wavelength emission is roughly proportional to the average 
number of UV absorption events that occur before a re-emitted photon 
escapes.  Thus a higher UV optical depth implies relatively more 
photons with $\lambda > 4000$ {\AA}.  Consequently the 
violet-to-red emission lines strongly depend on the \ion{Fe}{2} density, 
unlike the UV lines.  This is why 
the features noted by \citet{me10b} weakened dramatically without 
a proportionate change in the UV spectrum. 

    \vspace{1mm}   

\subsection{The UV brightening}   
\label{subsec:contin_b}

The flux levels in our UV data increased by factors of about 5 between  
2002 and 2015, see below.  Since $\eta$ Car is fairly close to the Eddington 
Limit, it cannot brighten intrinsically by such a factor.  Hence the observed 
brightening must indicate a rapidly declining circumstellar extinction.  
Two complications 
must be acknowledged, however.  First, what we call ``extinction'' may 
involve a scattering halo at $r \sim $ 100 to 500 AU \citep{hi06}, 
not just line-of-sight obscuration.  The halo is correlated with 
the circumstellar outflow density.  Secondly, the star's SED may have 
changed by a small amount.  In order to simplify the narrative, here we 
report the observed trend as though it were simple line-of-sight extinction 
by dust; future theoretical studies can make suitable corrections. 

    \vspace{1mm} 

\begin{deluxetable}{cccccc}   
    \tablenum{2}
\tablecaption{Quasi-continuum Flux Levels\tablenotemark{a}}  
\label{tab:uvflux}  
\tablehead{ 
\colhead{Date} & \colhead{1733\AA} & \colhead{1777\AA} 
	& \colhead{1968\AA} & \colhead{2072\AA} 
	& \colhead{2000-2200{\AA}\tablenotemark{b} }  }     
\startdata
2002.51  &  \ 3.2  & \ 3.2  & \ 2.1  & \ 2.0  &  1.80  \\    
2004.18  &  \ 2.9  & \ 3.5  & \ 2.8  & \ 2.5  &  1.90  \\  
2013.82  &   13.0  &  12.8  & \ 8.9  & \ 7.5  &  6.48  \\   
2015.67  &   16.8  &  17.0  &  12.2  &  11.3  &  8.49  \\
\enddata  
\tablenotetext{a}{$f_\lambda$ expressed in units of $10^{-12}$ 
   erg cm$^{-2}$ s$^{-1}$ {\AA}$^{-1}$, not corrected for extinction. 
   The four narrow samples are each 4 to 7 {\AA} wide.} 
   \tablenotetext{b}{The broad 2000-2200 {\AA} interval includes 
   many absorption lines.}
\end{deluxetable}  

    \vspace{1mm} 

 \begin{figure}   
 \plotone{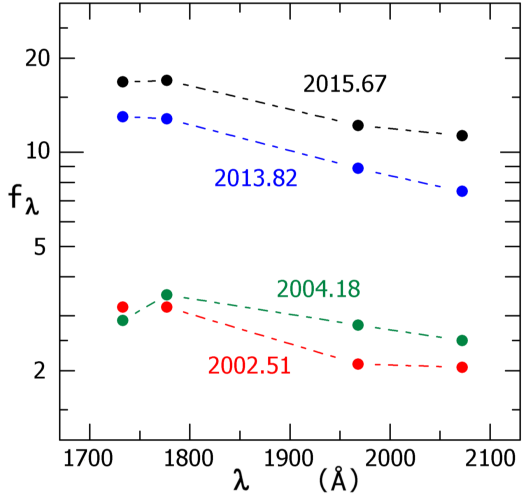}  
 \caption{Observed flux at four quasi-continuum wavelengths listed in   
    Table \ref{tab:uvflux}, not corrected for extinction.  Here the unit 
    for $f_\lambda$ is $10^{-12}$ erg cm$^{-2}$ s$^{-1}$ {\AA}$^{-1}$. }  
 \label{fig:contin} 
 \end{figure}

As quasi-continuum samples, we adopt four narrow wavelength intervals 
which appear as local plateaus in $f_\lambda$;  see Table \ref{tab:uvflux}.  
These provide meaningful flux     
measures even if they do not represent a true continuum, because  
(1) they had similar plateau-like appearances at each of the observation 
dates, and (2) their relative flux ratios remained consistent 
through the 2002, 2004, 2013, and 2015 observations.   
These $f_\lambda$ samples are plotted in Figure \ref{fig:contin}.  
Table \ref{tab:uvflux} and most of our figures do not 
include corrections for extinction, because $A_\lambda$ is only  
vaguely known for this object.  A primary goal of the next few 
paragraphs is to estimate the extinction.

   \vspace{1mm}  

The uncertainties in Table \ref{tab:uvflux} are not statistical, but 
likely error sizes can be be estimated with a few assumptions.  Suppose 
that the intrinsic fluxes at 1733, 1777, 1968, and 2072 {\AA} 
were constants, and that changes of extinction had the local wavelength  
dependence ${\Delta}A_\lambda \propto {\lambda}^{\alpha}$ in the range  
1700-2100 {\AA},  with a constant value of $\alpha$.  Given the 
16 data points, one can calculate a best-fit model that has eight parameters:  
three numbers that describe the intrinsic fluxes relative to each other,  
values of ${\Delta}A_{\lambda}(2000 \; \mathrm{\AA})$ at four observation 
times, and $\alpha$.  (The best-fit $\alpha$ turns out to be about $-0.5$.) 
Then we can deduce typical errors  from differences between the measured 
$f_\lambda$ values and the model, allowing for its degrees 
of freedom.  Based on this method, the r.m.s. error in each listed 
$f_\lambda$ value is roughly $\pm 10$\% or $\pm 0.11$ magnitude, much 
larger than the statistical count-rate errors.   
This probably involves real fluctuations in the spectrum,  and the 
largest deviations occurred at the shortest wavelength 1733 {\AA}.     

   \vspace{1mm}  
 \begin{figure}   
 \plotone{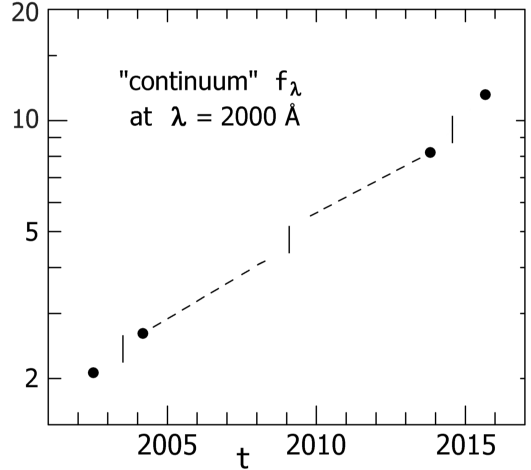}  
 \caption{ Progressive increase in apparent flux, an average of the 1968 
    {\AA} and 2072 {\AA} data listed in Table \ref{tab:uvflux}.  The 
    unit of $f_\lambda$ is $10^{-12}$ erg cm$^{-2}$ s$^{-1}$ {\AA}$^{-1}$.
    Vertical marks indicate the times of periastron events in 2003, 
    2009, and 2014.  A conjectural discontinuity in 2009 is consistent 
    with HST photometry at longer wavelengths (Martin et al.\ 2018, 
    in preparation).   }  
 \label{fig:brighten} 
 \end{figure}

As noted above, the five-fold 
increase in Figure  \ref{fig:contin} represents primarily  
a decrease in extinction.  Most likely the rate of mass outflow 
along our line of sight has diminished, so less dust is being formed 
at $r \sim $ 100 to 400 AU \citep{mar06b,mar10,hm12,da99b}.
Figure  \ref{fig:brighten} shows the trend for an average of the 
fluxes centered at $\lambda \approx $ 1968  and 2072 {\AA}.   The time-averaged  
rate of brightening from 2002 to 2015 was between 12\% and 16\%  per year 
at all four wavelengths in Table \ref{tab:uvflux}, with an overall average 
of 13.8\% per year or 0.14 $\pm$ 0.01 magnitude y$^{-1}$.   
(Here the quoted error may be doubtful because it is based on 
the $f_\lambda$ uncertainty estimated above, where normal statistics 
do not apply.) 
The brightening rate was approximately proportional to ${\lambda}^{-0.5}$ 
in the sampled range 1733--2072 {\AA}.  
``Normal'' extinction increases with $\lambda$ in this range 
\citep{ca89}, but $\eta$ Car's extinction law is notoriously peculiar  
with large grains and nitrogen-rich gas, see refs.\ in  
\citet{wa12} and \citet{dh97}.  Moreover, the ``UV halo'' effects 
mentioned earlier may affect the wavelength dependence. 
These remarks apply only to recently formed outflowing dust 
within 1000 AU of the star.  

  \vspace{1mm} 

Other wavelength samples give similar rates of change.    
For instance, the 2000-2200 {\AA} flux in Table \ref{tab:uvflux} 
is a simple average across that broad interval, including many 
spectral lines.  It shows an average brightening rate of 0.13 
magnitude per year. 

   \vspace{1mm}

We can estimate the total UV extinction, based on the following 
assessment of intrinsic brightness.  Since $\eta$ Car's 
photospheric temperature is probably in the range 15000--25000 K 
\citep{da12}, most of its emergent radiation should be in the wavelength 
range 1000--3000 {\AA}.  Scattering effects noted by 
\citet{hi06} cannot  shift a major fraction of the integrated 
flux out of this range.  For simplicity, first   
consider a Planck distribution with a Lyman cutoff near 912 {\AA}.   
If $\Phi_{2000}$ denotes the intrinsic value of $\lambda f_\lambda$ 
at  $\lambda = 2000$ {\AA} (i.e., what we would see in the 
absence of extinction), and $F$ is the total intrinsic energy flux 
$\int f_\lambda d\lambda$ including all wavelengths,  then  
$0.5 F < \Phi_{2000} < 0.8 F$   for any Planck temperature in the range 
noted above.  Further  details are too lengthy to explore here, 
but $\Phi_{2000} \approx 0.6 F$ appears probable, within a factor 
of 1.6 or so, for a realistic SED with free-free emission, a scattered 
halo, and other effects.  Given the standard 
luminosity $L \approx 4 \times 10^6 \ L_\odot$ for the primary 
star and $D \approx 2.3$ kpc, we thus expect 
$\Phi_{2000} \sim 1.5 \times 10^{-5}$ erg cm$^{-2}$ s$^{-1}$ 
and $f_\lambda(2000 \; \mathrm{\AA}) \sim 7.5 \times 10^{-9}$ 
erg cm$^{-2}$ s$^{-1}$ {\AA}$^{-1}$ without extinction.  
A factor-of-two  error in this quantity would alter the extinction 
$A_\lambda$ derived below by only about 10\%.

  \vspace{1mm}  

In 2015 the apparent $f_{\lambda}(2000 \; \mathrm{\AA})$ was smaller 
than the above value by a factor of about 630 (Table \ref{tab:uvflux}), 
implying 7.0 magnitudes of extinction.  Since $\eta$ Car's 
{\it interstellar\/} extinction amounts to roughly 3.3 magn at 
2000 {\AA},\footnote{   
  $A_V \approx 1.5$ \citep{dh97} with an $R = 4$ Cardelli law 
  \citep{ca89}.  }  
the circumstellar extinction in 2015 was evidently about  
3.7 magn at 2000 {\AA} with an informal pseudo-sigma uncertainty of the 
order of $\pm$0.6 magn.  Thirteen years earlier the star  
appeared 1.8 magn fainter (Fig.\ 4),  
so the UV circumstellar extinction 
decreased by roughly 33\%.  Note, however, that an unknown     
fraction of the ``circumstellar'' extinction occurs in older ejecta 
located at $r \sim 10^3$ to $10^4$ AU, which would have changed  
at a much smaller rate due to expansion.  Hence, allowing for normal 
uncertainties, the current dust-formation  rate appears to have fallen 
by more than 30\% between 2002 and 2015.  Extrapolating back to 1998 
when $\eta$ Car's rapid brightening was first noticed, we conclude 
that {\it the dust-formation rate along our line of sight 
decreased by at least 35\%} and probably more, perhaps    
about 50\%.   

   \vspace{1mm}

\subsection{Related issues}  
\label{subsec:contin_c}

Part of the measured brightening may involve a decrease in the size 
of a UV scattering halo \citep{hi06};  but that too would  
indicate a diminished outflow density, so the basic implication is 
not materially altered.   (The STIS/MAMA 0.2-arcsec aperture included 
most of the halo according to Hillier's Figure 15.)  One might attribute 
the trend to a destruction of dust grains, rather than a decreased rate 
of grain    formation;  but  that would require a change 
in the UV output or some other characteristic of the star, which would 
alter the mass flow rate as well.  As \citet{da99b} explained, the 
observed trend has been far too rapid to be a mere consequence of 
expansion or sideways motion of pre-existing dust.  That paper also noted 
reasons to suspect that the effect is less dramatic in some other 
directions from the star.    Incidentally,  outflowing gas reaches 
the dust-formation region about two years after leaving the star. 

    \vspace{1mm}  

In principle the temporal baseline can be extended about a decade earlier, 
because $\eta$ Car was observed in 1991--1997 with HST's Faint Object 
Spectrograph (FOS) and Goddard High Resolution Spectrograph (GHRS)  
\citep{da95,hu99,eb97}.  One datum is highly relevant here:   
at $t = 1991.62$, near the same orbital phase as 2002.51 and 2013.82, 
the average apparent $f_\lambda$(1450-1700 {\AA}) was {\it probably\/} 
about $7 \times 10^{-13}$ erg cm$^{-2}$ s$^{-1}$ {\AA}$^{-1}$ \citep{da95}.  
We emphasize ``probably'' because those early FOS data required 
unorthodox analysis to achieve sufficient spatial resolution.  
Relative to Table \ref{tab:uvflux}, {\it one deduces a UV brightening 
rate of roughly 0.13 magn y$^{-1}$ from 1991 to 2002\/} -- practically the  
same as in 2002--2015.  

    \vspace{1mm}  

This result is moderately surprising for two reasons: (1) the accelerated 
brightening in ground-based photometry began several years after 1991;  
and (2) in the early 1980's, the star appeared definitely brighter than 
the Weigelt knots \citep{we86}.  Since their energy budget and line ratios 
indicate much less extinction than for the star 
\citep{ha12,da97,dh97}, a naive extrapolation of the 1998-2015 trend 
back to 1984 would make the star no brighter than the knots at that 
time, even at far-red wavelengths.   
{\it The FOS and GHRS data, and also the International Ultraviolet 
Explorer (IUE) observations of $\eta$ Car in the 1980's, merit 
new examinations relative to the STIS data\/} -- a task beyond 
the scope of this paper. 

   \vspace{1mm}  

Concerning the FUV quasi-continuum level (Fig.\ 2),  
one must be careful with the word ``photosphere'' in a diffuse flow. 
In order to be physically meaningful as well as consistent with 
traditional usage, a photosphere should be the region that determines 
the emergent photon energy distribution.   In a hot atmosphere  
or wind,  this is the thermalization depth where 
$ (3 \, \tau_\mathrm{tot} \, \tau_\mathrm{abs})^{1/2} \, \approx \, 1$.  
If $\eta$ Car has $\dot{M} \gtrsim 3 \times 10^{-4} \ M_\odot$ y$^{-1}$,
then its photosphere defined this way is located in the wind rather than 
near the star's surface \citep{da87}.\footnote{ 
   Calculations by \citet{ow16} agree in their essentials with 
   the simplified view in \cite{da87}, if we allow for modern 
   opacity values.  Their discussion seems to imply otherwise, 
   largely because they quoted only textual comments rather than the 
   quantitative temperatures shown in Davidson's Figure 1. }  
The single-star wind model described by \citet{hi01}, for example, had  
a characteristic photosphere temperature somewhat above 15000 K.   
The classical effective temperature $T_\mathrm{eff}$ has no physical 
significance in a diffuse configuration.   

   \vspace{2mm}

\begin{deluxetable}{cccc}   
\tablenum{3}
\tablecaption{The \ion{N}{3}] $\lambda$1750 Multiplet\tablenotemark{a} } 
\label{tab:1750}    
\tablehead{ 
\colhead{${\lambda}_0$ (\AA)} 
   & \colhead{$J_\mathrm{lower}$\tablenotemark{b} } 
   & \colhead{$J_\mathrm{upper}$\tablenotemark{b} }   
   & \colhead{ \ \ $A_{ji}$ (s$^{-1}$) }  }    
\startdata 
    1746.823  &  1/2  &  3/2  & \ \   8.8  \\  
    1748.646  &  1/2  &  1/2  &     346.8  \\  
    1749.674  &  3/2  &  5/2  &     266.0  \\  
    1752.160  &  3/2  &  3/2  &  \   60.2  \\   
    1753.995  &  3/2  &  1/2  &     361.5  \\  
\enddata  
\tablenotetext{a}{See \citet{be95} and \citet{st94}.}
\tablenotetext{b}{Lower and upper terms are 
   $2s^{2}2p \ ^{2}\mathrm{P}^\mathrm{o}$ and  
   $2s2p^{2} \ ^{4}\mathrm{P}$ respectively. }     
\end{deluxetable}

\section{Extremely Luminous \ion{N}{3}] Emission} 
\label{sec:niii1750}

The secondary star in $\eta$ Car has $T_\mathrm{eff} \sim  40000$ K 
\citep{me10a}, hot enough to photoionize helium in some parts of the 
primary star's wind.  Nitrogen is more 
abundant there than carbon plus oxygen 
\citep{da86}, and tends to be doubly ionized in a zone of He$^+$.     
Therefore semi-forbidden \ion{N}{3}] $\lambda$1750 emission 
(Table \ref{tab:1750}) is a very strong coolant in the He$^+$  
zone.  Three attributes together make this feature unique in 
$\eta$ Car's spectrum:  (1) It is fundamentally strong as just 
noted;  (2) it has a rather high excitation energy;  and 
(3) being semi-forbidden, it has no P Cyg absorption or other 
self-absorption.  

    \vspace{1mm}

The \ion{N}{3}] multiplet provides evidence for a long-standing 
question,  the location of the photoionized He$^+$ region 
within the wind structure. It also reveals some fresh problems.   
Throughout the following discussion, we invoke numerous conventional 
parameters for the $\eta$ Car system.  For explanations of them and 
references, see the review articles in \citet{dh12}.

    \vspace{2mm}

 \begin{figure}   
 \plotone{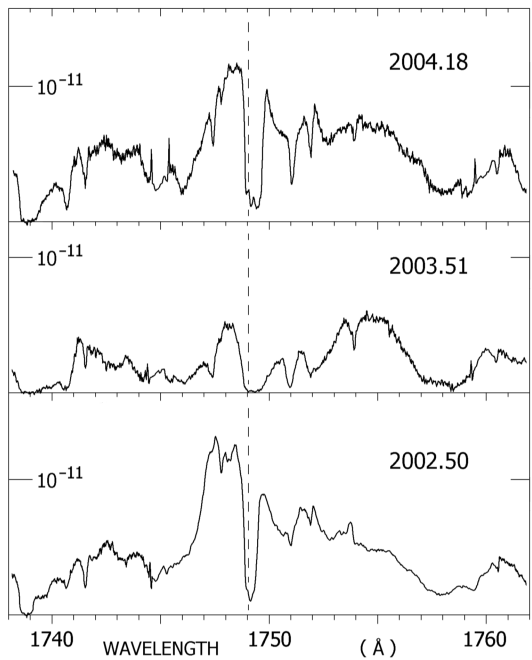} 
 \caption{\ion{N}{3}] $\lambda$1750 emission observed in 2002-2004. 
   Horizontal marks indicate 
   $f_{\lambda} = 10^{-11}$ erg cm$^{-2}$ s$^{-1}$ {\AA}$^{-1}$, not 
   corrected for extinction.  A vertical dashed line  
   marks $-500$ km s$^{-1}$ for the \ion{Ni}{2} $\lambda$1751.9 
   absorption line.    
   The middle panel is atypical because it represents 
   a brief  ``spectroscopic event'' at periastron. }
 \label{fig:niii2002} 
 \end{figure}
 \begin{figure}   
 \plotone{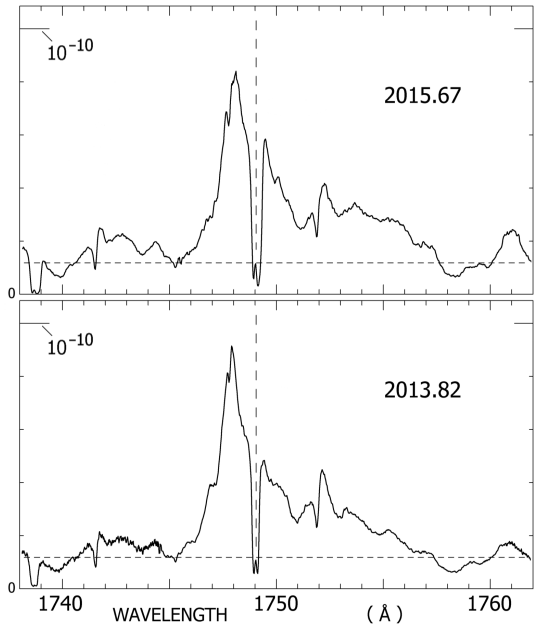} 
 \caption{\ion{N}{3}] $\lambda$1750 emission in 2013-2015.  
   Horizontal marks indicate 
   $f_\lambda = 10^{-10}$ erg cm$^{-2}$ s$^{-1}$ {\AA}$^{-1}$,
   not corrected for extinction.  A vertical dashed line  
   marks $-500$ km s$^{-1}$ for the \ion{Ni}{2} $\lambda$1751.9 
   absorption line.  } 
 \label{fig:niii2013} 
 \end{figure}
\subsection{ The \ion{N}{3}] luminosity }
\label{subsec:niiia}

   \vspace{1mm} 

\ion{N}{3}] emission, like the \ion{He}{1} recombination lines, does 
{\it not\/} depend primarily on the amount of material present. 
Instead it represents the EUV luminosity of the hot secondary star.  
Most photons with $h\nu \gtrsim 25$ eV are absorbed by  
He$^0$ $\rightarrow$ He$^+$ events somewhere in the gas flow,   
and most of their  energy flux is then recycled to 
$h\nu \lesssim 10$ eV by recombination, \ion{N}{3}] emission, 
and other cooling processes \citep{hu08}.    
To a first approximation, the \ion{N}{3}] luminosity is thus 
independent of the gas density distribution.  
The line profile, however, depends on flow velocities of    
the EUV-absorbing material.   
Figures \ref{fig:niii2002} and \ref{fig:niii2013} show the observed profile 
of \ion{N}{3}] $\lambda$1750 in 2002-2015.  The very different-looking 
middle panel in Figure \ref{fig:niii2002} represents the 2003.5 periastron 
event which was too complicated to discuss here 
\citep{dh12,mar06a,me11,me15}.  

    \vspace{1mm} 

The luminosity of \ion{N}{3}] emission can be estimated as follows. 
For a reason noted in {\S}\ref{subsec:niiie}, let us focus 
on the data at $t = 2013.82$.  The integrated apparent flux was then     
$F$(\ion{N}{3}]) $\approx \ 2.4 \times 10^{-10}$ erg cm$^{-2}$ s$^{-1}$, 
continuum subtracted and not corrected for extinction.  
Here we interpolated across the absorption lines via a model in 
{\S}\ref{subsec:niiib} below, and the informal pseudo-sigma uncertainty 
is roughly $\pm$15\%.  Extinction at 1750 {\AA} amounted to 7.7 $\pm$ 0.7 
magnitudes ({\S}\ref{subsec:contin_b} above, adapted to 2013 
rather than 2015).  Hence the flux without extinction would be 
$F_0$(\ion{N}{3}]) $\approx 3 \times 10^{-7}$ erg cm$^{-2}$ s$^{-1}$. 
This implies luminosity 
$L$(\ion{N}{3}]) $\approx \ 2 \times \ 10^{38}$ ergs s$^{-1}$ 
$ \approx \ 5 \times 10^4 \; L_\odot$, 
which is enormous for a stellar-wind emission feature.   
The uncertainty is a factor of about 2.  

   \vspace{1mm}

This value exceeds the total kinetic energy outflow of the wind,
and a single-star wind produces far less \ion{N}{3}] 
emission \citep{hi01,hi06}.  Hence this feature is almost certainly 
powered by ionizing radiation from the hot secondary star, perhaps 
supplemented by an effect noted in {\S}6 of \citet{hu08}.  But the 
energy budget seems problematic in light of the following facts: 
\begin{enumerate} 
    \item The 40000 K companion star discussed by \citet{me10a} 
    radiates less than $10^5 \; L_\odot$ at helium-ionizing photon 
    energies $h\nu > 24.6$ eV. 
  \item Even though \ion{N}{3}] $\lambda$1750 emission may be the 
    strongest individual cooling mechanism in a He$^{+}$ zones,  
    other processes should account for much of the total  cooling.  
  \item Collisional de-excitation reduces the efficiency 
    of the $\lambda$1750 emission ({\S}\ref{subsec:niiib} below). 
  \item In any likely geometry ({\S}\ref{subsec:niiic} below), some  
    fraction of the secondary star's radiation escapes 
    along paths through the secondary wind, which is not dense enough 
    for appreciable absorption or emission. 
\end{enumerate} 
Hence the observed feature is brighter than we would have predicted.   
Several possible explanations are available.   We may have overcorrected 
$F$(\ion{N}{3}]) for extinction; or  the secondary star may be hotter 
than 40000 K; or FUV photons from the primary star may contribute 
to the heating via a trick noted in {\S}6 of \citet{hu08}; etc.  The 
main point is that $L$(\ion{N}{3}] $\lambda$1750) {\it appears to be 
comparable to the maximum  attainable value.} 

    \vspace{4mm}  
 \begin{figure}   
 \plotone{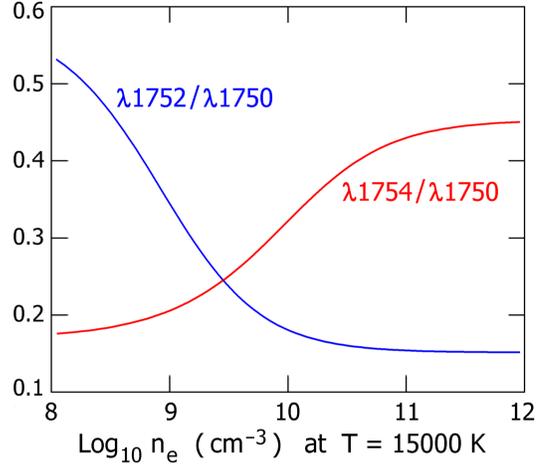}
 \caption{Density-dependent line ratios in the \ion{N}{3}] multiplet. } 
 \label{fig:niiiratios} 
 \end{figure}

\subsection{Density of the He$^+$, N$^{++}$ zone}     
\label{subsec:niiib}  

Gas density $n_e$ can help to indicate the location of the He$^+$, 
N$^{++}$ region, as explained later.  
Relative intensities of the \ion{N}{3}] multiplet components 
(Table \ref{tab:1750}) depend approximately on $\, n_e/T^{0.4}$.  
We calculated them across a broad density range, 
using collision strengths and radiative rates reported by \citet{st94} 
and \citet{be95}.  Let us adopt 
$T = 15000$ K;  for other temperatures one can simply multiply each 
quoted $n_e$  value by $(T/15000 \; \mathrm{K})^{0.4}$.  Figure 
\ref{fig:niiiratios} shows the brightness ratios of the three most 
useful lines $\lambda$1750, $\lambda$1752, $\lambda$1754 as functions 
of density. 

   \vspace{1mm}

Consider the HST data at $t = 2013.82$, the lower panel 
in Figure \ref{fig:niii2013}. 
Strong peaks near 1748 and 1752 {\AA} represent the $\lambda$1750 
and $\lambda$1754 lines shifted by $-300$ km s$^{-1}$, 
and another member of the multiplet can be discerned near 
1747 {\AA}.  But there is no peak at 1750 {\AA} corresponding to  
the $\lambda$1752 line. Evidently $\lambda$1752 is 
considerably fainter than $\lambda$1754, so Figure \ref{fig:niiiratios} 
immediately indicates $n_e \gtrsim 10^{10}$ cm$^{-3}$.     

  \vspace{1mm}

\begin{figure}   
 \plotone{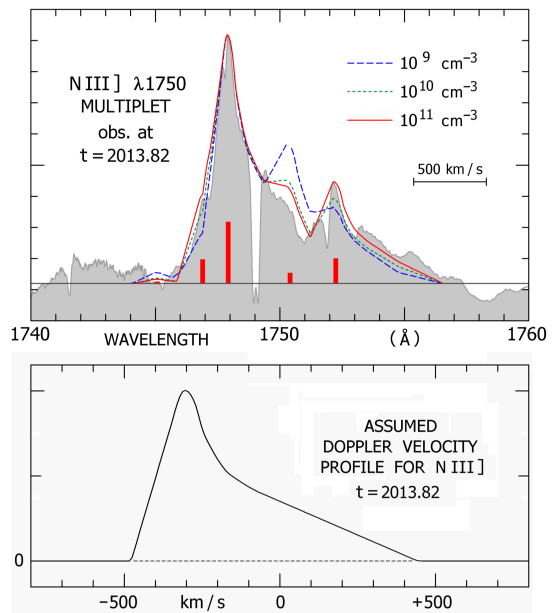}  
 \caption{Observed \ion{N}{3}] $\lambda$1750 profile (shaded), compared 
    to models for densities $n_e  = 10^9, \; 10^{10}, \; 10^{11}$ 
    cm$^{-3}$  (see text).  Small vertical bars indicate relative 
    strengths of the multiplet components in the high-density limit.  
    Each model used the Doppler profile shown in the bottom panel. }
 \label{fig:niiiprofile} 
 \end{figure}

In order to be more definite, we need a model for the emission profile.   
Unrelated absorption lines prevent us from using a deconvolution 
technique to separate the multiplet structure from the underlying 
Doppler profile.  Instead we estimated the latter by informal 
trial-and-error experiments.  The adopted Doppler profile is shown in the 
lower panel of Figure \ref{fig:niiiprofile}.  We are not confident 
of the longer-wavelength tail extending to $+450$ km s$^{-1}$, 
but it fits the data reasonably well and its influence on the density 
estimate is smaller than other uncertainties.  The upper panel 
in  Figure \ref{fig:niiiprofile} shows the resulting total profile 
for $n_e = 10^{9}, \ 10^{10}$, and $10^{11}$ cm$^{-3}$, normalized 
to the major peak.   A density of $10^9$ cm$^{-3}$ is clearly 
unsuitable,  $10^{11}$ cm$^{-3}$ gives the best fit, and higher 
densities are practically indistinguishable from $10^{11}$ cm$^{-3}$. 
The main uncertainty results from irregularities in the underlying 
fluxes at 1752 {\AA} vs.\ 1748 {\AA} -- i.e., the subtracted 
``continuum.''  Assuming that this 
wavelength region has the same statistical distribution of relative 
fluxes as the 2000-2200 {\AA} interval (Fig.\ \ref{fig:uvtracing}), 
we estimate that the probabilities of $n_e < 10^{10.0}$ cm$^{-3}$ and 
$n_e < 10^{10.3}$ cm$^{-3}$ are less than 5\% and 22\% respectively. 
A more elaborate analysis, employing a model of the superimposed 
\ion{Fe}{2} and \ion{Ni}{2} spectrum, would reduce the uncertainty 
but is far too lengthy to attempt here. 

    \vspace{1mm} 

Collisional de-excitation \citep{of06} reduces the efficiency of 
\ion{N}{3}] $\lambda$1750 in the favored density range, because  
its critical density is $n_c \approx 10^{10.4}$ cm$^{-3}$.  
Note, however, that \ion{N}{3}] emission is a dominant  
cooling process and it depends strongly on temperature.    
If collisional de-excitation reduces the emission efficiency, then 
the equilibrium temperature rises so \ion{N}{3}] is almost as strong 
as it would have been without collisional de-excitation.  
If nearly 
all of the cooling is due to this feature plus recombination and 
free-free emission, then the resulting \ion{N}{3}] luminosity is 
reduced by 25\% at $n_e \sim 10^{11.2}$ cm$^{-3}$, and by 50\% 
at $10^{11.7}$  cm$^{-3}$.  Expected temperatures are in the range 
13000-23000 K depending on density.    


   \vspace{1mm} 

In summary, the \ion{N}{3}] $\lambda$1750 data favor the density 
range  $10^{10.3} \lesssim n_e \lesssim 10^{11.5}$ cm$^{-3}$.  
Lower values produce 
unsatisfactory multiplet ratios, while higher densities entail excessive 
collisional de-excitation.  The dominant uncertainties are not 
statistical and they are not caused by imperfections in the data;  
instead they involve the complex nature of 
$\eta$ Car's spectrum.  This estimate applies to gas with 
Doppler velocities near $-300$ km s$^{-1}$, because it is based essentially 
on the peaks seen near 1748 and 1752 {\AA}.  Also, of course, it presumably 
represents a weighted average for gas that has a range of densities.

   \vspace{1mm}

\subsection{Location of the N$^{++}$, and small-scale structure in 
   the shocked region}     
\label{subsec:niiic}  

   \vspace{1mm} 

\begin{figure}   
 \plotone{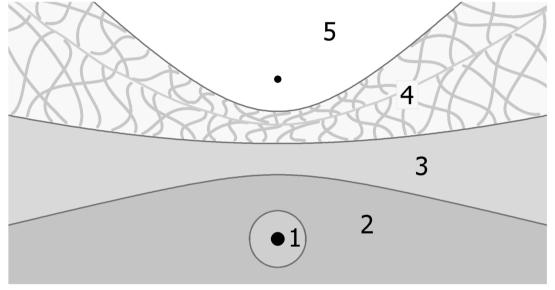} 
 \caption{Conceptual arrangement of ionization zones in the two stellar 
   winds.  The primary and secondary stars are indicated near the botton 
   and top respectively.  The primary wind includes zones 1, 2, and 3, 
   while the less dense secondary wind is zone 5.  Region 4 is the complex 
   shocked region.  This sketch is {\it highly\/} idealized, because 
   every zone is unstable and inhomogeneous. } 
 \label{fig:zones} 
 \end{figure}

Where is this gas located in $\eta$ Car's wind structure?  Figure  
\ref{fig:zones} identifies five different zones, with the primary 
star at the bottom and the secondary star above it.  Every part of 
this idealized map is inhomogeneous and unstable, but  the zones   
are meaningful regarding ionization states. 
Regions \underline{1,2,3} are differently ionized parts of the 
primary wind, which has $v \sim 500$ km s$^{-1}$.   Zone \underline{4} 
is a shocked region between the two winds, and \underline{5} is the 
lower-density secondary wind with $v \sim 3000$ km s$^{-1}$.  
He$^+$ and N$^{++}$ may exist in each zone except \underline{2}.  
We tentatively dismiss the inner wind \underline{1} in this problem, 
because relevant single-star wind models do not produce much 
\ion{N}{3}] emission \citep{hi01,hi06}; but this statement is not 
absolutely robust and zone \underline{1} will be mentioned again in 
{\S}\ref{subsec:niiid}.  The secondary wind \underline{5} has far 
too low a density, so we are left with zones \underline{3} and 
\underline{4}.

    \vspace{1mm}  

Zone \underline{3} is the part of the primary wind where helium 
may be photoionized by the hot secondary star.  If the wind had  
$\dot{M} \approx 10^{-3} \; M_\odot$ y$^{-1}$ and 
$v \approx 500$ km s$^{-1}$  (see reviews in \citealt{dh12}), 
then the average density there at $t = 2013.82$ would have been 
$n_e \sim 10^{9.2}$ cm$^{-3}$, far below the range indicated by the 
\ion{N}{3}] multiplet ratios.  Instabilities, however,  
cause the wind to be inhomogeneous, perhaps with localized density 
maxima above $10^{10}$ cm$^{-3}$ \citep{hi01},  and 
emission lines originate chiefly in those high-density locales.  
On the other hand, the mass loss rate probably declined well below  
$10^{-3} \ M_\odot$ y$^{-1}$ before 2013, if our interpretation 
of the secular changes is even partially valid.    
In summary,  region \underline{3} is {\it probably\/} not dense 
enough,  but a resourceful skeptic can devise models that avoid 
this result.  The existence or non-existence of this He$^+$ zone      
is a non-trivial question, see below.  

    \vspace{1mm}  

The shocked region \underline{4} is hostile to simple analysis, and 
below we shall note reasons to mistrust the numerical simulations that 
have appeared so far.   Boundary \underline{3-4} is not a 
well-defined shock front, because it is extremely unstable. 
Consider a sample  of gas flowing from region \underline{3} into 
\underline{4}.  After being compressed by a factor of about 4 and 
heated to $T \sim 10^{6.6}$ K,  
{\it it is thermally unstable\/} with a cooling time less than a day
\citep{dr11}.  When the resulting small cloudlets or filaments have 
cooled to some level below 30000 K, their temperatures stabilize 
because of UV heating   by the two 
stars.  (Here ``cloudlet'' means, essentially, a local density 
maximum.)   Meanwhile, gas from the secondary 
wind forms a hot medium in region \underline{4}, because 
its shock front \underline{4-5} has  $T \gtrsim 10^8$ K with a cooling 
time of the order of 0.3 year.  Judging from the sound speeds and 
timescales,  pressure equilibrium has a rough validity.  Based on 
pressure quasi-equilibrium, the cooled cloudlets should have 
densities $n_e \sim 10^{11}$ to $10^{12}$ cm$^{-3}$.  This picture 
is obviously consistent with our \ion{N}{3}] results.   

    \vspace{1mm}  

So far as we know, the morphology of the cooled shocked gas has 
not yet been modeled with a realistic range of size scales and 
physical processes.  The mass and size of a typical cloudlet may 
result from one or the other of two effects:   
(a) Each ``clump'' in the inhomogeneous primary wind might become a 
cloudlet, or (b) alternatively, the  characteristic size scale for thermal 
instability may be more decisive: $\ell \sim$ (sound speed) $\times$ 
(cooling time) $\sim$ 0.03 AU.\footnote{
   A classical adiabatic shock would produce 
   $T \approx 4 \times 10^6$ K there, implying a sound speed of about   
   200 km s$^{-1}$ and cooling time of $2 \times 10^4$ s or less.   
   Instabilities, however, may cause the transition \underline{3-4}  
   to be a succession of oblique subshocks.  In that case the 
   resulting temperature is lower and the cooling time is shorter. }
This suggests a size less than 0.01 AU for a typical cloudlet or filament 
after it has shrunk due to cooling and surrounding pressure.  
These condensations might be rapidly disrupted by Kelvin-Helmholtz 
and/or Rayleigh-Taylor effects, photoionization-driven evaporation, etc., 
but the observed strong \ion{N}{3}] emission with  
$n_e \sim 10^{11}$ cm$^{-3}$ suggests that this is not the case.  

    \vspace{1mm} 

{\it Do they absorb nearly all helium-ionizing photons that enter the 
shocked region?\/}  If not, then a He$^+$ zone \underline{3} exists in 
the primary wind.   Each cloudlet is opaque at $\, h\nu \sim 25$ to 
30 eV, but photons might pass between them.  Therefore, as \citet{me12} 
emphasized, the answer depends on sizes, shapes, orientations, and 
geometrical correlations of the cloudlets.  (See the last part of that 
paper's  {\S}3.1.)  Imagine  a conceptual model with average cloudlet 
size $s \approx 0.007$ AU.  A density compression factor of 100 in 
each cloudlet implies volume filling factor $\epsilon \approx 0.01$, so 
the number density of cloudlets would be 
$N \sim {\epsilon}/s^3 \sim 10^{4.5}$ AU$^{-3}$.
In order to block almost all photon paths through the shocked region, 
the mist of cloudlets must extend to a thickness 
$H \; \gtrsim \; 1/s^2 N \sim 0.6$ AU 
perpendicular to boundary \underline{3-4}.  Given the flow speeds in 
that region,  this extent requires a cloudlet survival 
time of at least 10 days,  which is roughly 10 times the sound-speed 
crossing time in an individual cloudlet.  It is difficult to say 
whether this survival time is theoretically reasonable.

    \vspace{1mm} 

\citet{cl14,cl15} described elaborate numerical simulations of 
the shocked flow, but they did not clarify the small-scale morphology.  
Roughly $10^6$ cloudlets are required, so a valid ``global'' 
simulation needs at least $10^8$ adaptive sample points with 
small local time steps -- far more than the Clementel et al.\/ 
figures appear to indicate.  Those authors' density maps show maxima 
below $10^{10.3}$ cm$^{-3}$.  Perhaps these are averages over 
regions large enough to portray in a figure, but {\it the root of 
the problem lies at small size scales} $\, {\Delta}x \lesssim 0.005$ AU,
which are not shown.  Moreover, the large compression factors suggest 
that MHD effects should be included.  If some factor prevents 
compression to the small sizes suggested above,  then \ion{N}{3}] 
should have indicated lower densities.   Most helium-ionizing photons 
are probably absorbed in the shocked region as Mehner et al.\/ 
indicated, but this opinion is not based on numerical simulations, 
and it has not been proven by them.

\subsection{The puzzling Doppler profile} 
\label{subsec:niiid}  

Our best-fit distribution of Doppler velocities for \ion{N}{3}] 
(lower panel in Fig. \ref{fig:niiiprofile}) does not resemble  
simple models.  Based on the discussion above, one would expect  
the N$^{++}$ to flow roughly along pseudo-hyperboloidal 
surfaces like those sketched in Figure \ref{fig:zones} -- 
somewhat analogous to water on an umbrella. 
At the times of observation, the angle between our line of 
sight and the umbrella axis was probably in the range 
40$^\circ$ to 75$^\circ$, with us on the concave side.\footnote{  
   Quantitative details in this paragraph are based on 
   elementary geometry calculations that are too lengthy 
   and distracting to recite here.  They are  consistent 
   with most published models of the wind-wind structure.  
   But if the orbit orientation is reversed \citep{ka16}, 
   then the above discussion has to be modified. } 
In that situation, the positive-velocity tail of the Doppler 
distribution makes no apparent sense.  Assuming that the peak 
at $-300$ km s$^{-1}$, represents the nearest parts of the 
umbrella, the extreme redshift on the other side should have 
been less than $+250$ km s$^{-1}$. 
Moreover, $\eta$ Car's \ion{He}{1} recombination lines 
do not have conspicuous long-wavelength wings.  

   \vspace{1mm}  

Admittedly we used an informal method to estimate the \ion{N}{3}] 
Doppler profile, but a glance at Figure \ref{fig:niii2013}  
strongly suggests that the long-wavelength side of the profile 
extends farther than the short-wavelength side.

   \vspace{1mm} 

We have no good solution to this problem, but here are three  
speculative ideas.   First, the receding N$^{++}$ might be in a 
distant part of zone \underline{3} in the primary wind.  Figure 5b in 
\citet{me12} shows that positive velocities may exist there, and the 
lower density is acceptable because our estimate $n_e \sim 10^{11}$   
cm$^{-3}$ really applies only to the peak of the Doppler profile. 

    \vspace{1mm} 

A second, less orthodox idea is that \ion{N}{3}] might originate 
in the inner wind (zone \underline{1}), contrary to published wind  
models.  The density there has the right order of magnitude, and the 
energy supply in that region is adequate.  The redshifted 
wing might even be caused by Thomson scattering, since the 
optical depth there is of order unity in some 
models \citep{hi01}.  A non-spherical   wind might 
achieve such results, in more or less the same vein 
as \citet{gr12a,gr12b,gr15}. 

   \vspace{1mm} 

As a third possibility, the long-wavelength wing might be the 
same as the feature noted in {\S}\ref{subsec:niiie} below.  

    \vspace{1mm}  

Arguably the  \ion{N}{3}] profile ``should have'' differed between  
the 2013 and 2015 observations, because the angle between our 
line of sight and the shock axis differed by about $20^\circ$ 
between those two occasions.   
At least one would expect the $-300$ km s$^{-1}$ velocity to 
differ measurably between those orientations.  In fact, however, 
the two profiles  in Figure \ref{fig:niii2013} are practically 
indistinguishable,  apart from the variable feature noted  
below.    

    \vspace{1mm}   

 \begin{figure}   
 \plotone{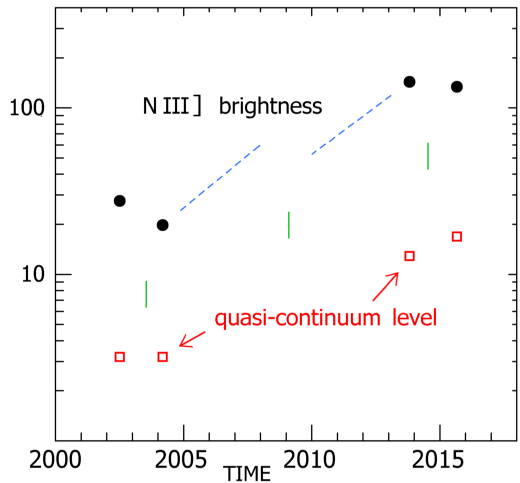} 
 \caption{Partial \ion{N}{3}] fluxes in range 1746.0--1749.0 {\AA},   
     expressed in units of $10^{-12}$ erg cm$^{-2}$ s$^{-1}$ with no 
     correction for extinction.  Open squares below are the averages 
     of $f_{\lambda}(1733)$ and $f_{\lambda}(1777)$ in Table \ref{tab:uvflux}, 
     in units of $10^{-12}$ erg cm$^{-2}$ {\AA}$^{-1}$.  Vertical marks  
     indicate periastron times.  A conjectural discontinuity in the 
     dashed trend line coincides with the 2009 periastron event,   
     cf.\/ Fig.\/  \ref{fig:brighten}.  } 
 \label{fig:1750vars} 
 \end{figure}

\subsection{Variability, and mysterious redshifted emission}    
\label{subsec:niiie}

In order to assess the \ion{N}{3}] brightness trend without too much 
absorption-line influence,  we integrated $f_\lambda$ 
across only the brightest, least affected interval  
1746.0--1749.0 {\AA}.  Figure \ref{fig:1750vars} shows the results    
with no extinction corrections nor continuum subtraction.  There 
are two noteworthy findings:   (1) The average rate of brightening 
was 0.17 magnitude per year, about 20\% faster than the nearby 
wavelengths listed in Table \ref{tab:uvflux}.  (2) In both 2002-2004 
and 2013-2015, the sample was noticeably fainter after 
periastron.  This deficit was comparable to the luminosity of the 
anomalous broad feature discussed next.  

    \vspace{1mm} 

The $t = 2004.18$ tracing in Figure \ref{fig:niii2002} shows {\it strong, 
broad emission around 1755 {\AA}, resembling a dominant emission bump seen 
during the 2003.5 periastron event.\/}  It also appears, 
less strongly, in the 2015.67 data  (Fig.\/ \ref{fig:niii2013}).  
In view of the innate strength of the \ion{N}{3}] multiplet, this 
feature is almost certainly the $\lambda$1750 line with  
Doppler velocities ranging from $+400$ to $+1200$ km s$^{-1}$.   
It had an impressive luminosity in 2003-2004,  
thousands of $L_\sun$. 
Considering Figures \ref{fig:orbit} and \ref{fig:zones} together, 
{\it we cannot easily find a locale for N$^{++}$ with 
these velocities.\/}   Certain parts of the secondary shock had   
large positive Doppler velocities in 2004, but the other parts 
would not,  and there is no reason to expect much N$^{++}$ 
near the secondary shock ({\S}\ref{subsec:niiic} above). 

    \vspace{1mm}  

Perhaps the 1755 {\AA} feature was a remnant of the 2003.5 periastron 
event that occurred 8 months earlier.  In the oldest, and in some 
respects the most successful interpretation of a periastron event in 
$\eta$ Car, $\dot{M}$ temporarily increases so a considerable amount of 
ejecta moves outward \citep{za84,mar06a,da12}.  Since the detailed 
mechanisms remain unclear, and the secondary star is probably located 
on the far side of the primary at periastron, we should not be 
surprised to see large recession velocities in the special ejecta  
-- for example,  the broad 1755 {\AA} emission in the middle panel 
of Figure \ref{fig:niii2002}.  Can it remain detectable later?  
By $t = 2014.18$ the special fast ejecta  would have been more than 
50 AU from the star, and N$^{++}$ would exist there only if there     
was a clear path for EUV photons from the secondary star to the 
receding material.  This may be possible due to the spiral   
patterns caused by orbital motion;  see Figures in  \citet{pa09} 
and \citet{ok08}, especially the right-hand panel in Figure 3 of 
Parkin et al.  However, this picture would not explain the same feature 
at 2002.51,  if it was weakly present then (Fig. \ref{fig:niii2002}). 

   \vspace{1mm}  

Evidently we have no satisfying explanation for the broad 
emission around 1755 {\AA}.  Of course the same can be said of 
some other spectral lines in $\eta$ Car; but this one has an    
extraordinary luminosity of the order of $10^4 \; L_\odot$.  
That amount is comparable to the total 
kinetic energy flow in the wind, and greatly exceeds the total 
X-ray flux.  As Oscar Wilde said, to misplace so much energy 
almost looks like carelessness.

\section{Other unusual emission features}    
\label{sec:emlines} 

  \vspace{1mm}  

Various authors have discussed the familiar types of UV stellar wind 
features seen in $\eta$ Car \citep{eb97,hi01,hi06,gr12a}.   
Most of those features have deep P Cyg absorption and require 
sophisticated analyses.  Here we note two special emission lines 
with different characteristics. 

   \vspace{1mm} 

\begin{figure}   
 \plotone{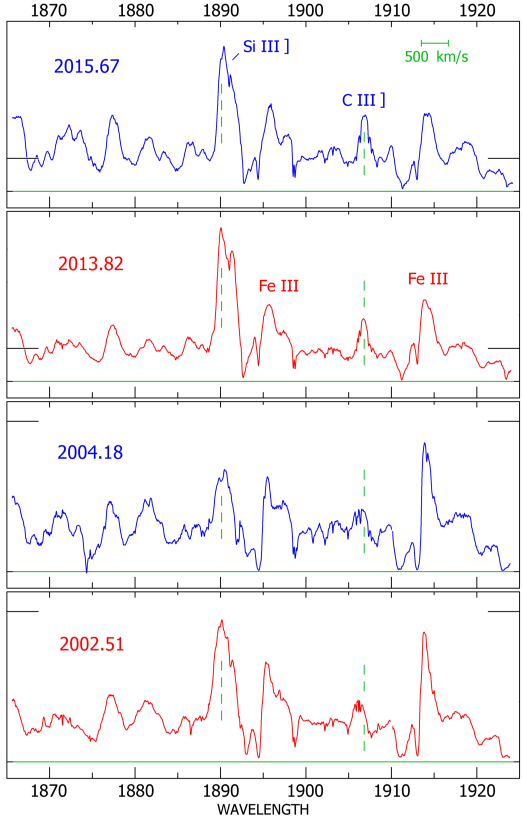} 
 \caption{Bright emission lines near 1900 {\AA} in $\eta$ Car's  
    spectrum.   Vertical dashed lines mark Doppler velocities of 
    $-300$ km s$^{-1}$ for \ion{Si}{3}] and \ion{C}{3}].  
    Horizontal marks at the sides indicate 
    $f_\lambda = 10^{-11}$ erg cm$^{-1}$ s$^{-1}$ {\AA}$^{-1}$,  
    not corrected for extinction. }  
 \label{fig:1900A}  
 \end{figure}

Figure \ref{fig:1900A} shows a grove of bright lines near  
$\lambda \sim 1900$ {\AA}, as seen in 2002, 2004, 2013, and 2015.  
Most of the unlabeled features are \ion{Fe}{2}, but the lines of 
interest here are semi-forbidden \ion{Si}{3}] and \ion{C}{3}] near 
1890 {\AA} and 1907 {\AA}.\footnote{  
  For a useful list of identifications in $\eta$ Car's spectrum, 
  see \citealt{vi89}. }  
Semi-forbidden lines are useful because their short-wavelength sides are 
not complicated by P Cyg absorption (see below).  From 2002 to 2015 the 
\ion{Si}{3}] and \ion{C}{3}] emission generally increased by modest 
amounts relative to \ion{Fe}{2} and \ion{Fe}{3}, and also became 
somewhat narrower.   Like the \ion{N}{3}] $\lambda$1750 peak,  
their average velocities were close to $-300$ km s$^{-1}$.    

    \vspace{1mm}  

\ion{C}{3}] $\lambda$1909 most likely originates in the same gas as 
\ion{N}{3}] $\lambda$1750, since the ionization 
energy for C$^+$ $\rightarrow$ C$^{++}$ is nearly the same as for 
He$^0$ $\rightarrow$ He$^+$.  Based only on the nature of the $\eta$ Car 
system, we expect the luminosities of \ion{N}{3}] and \ion{C}{3}] to 
depend chiefly on the flux of helium-ionizing photons from the secondary 
star.   However, the critical $n_e$ for collisional de-excitation of 
\ion{C}{3}] is   about $10^{9.7}$ cm$^{-3}$, far below the density range 
favored  in {\S}\ref{subsec:niiib} above.  
If $n_e \sim 10^{11}$ cm$^{-3}$, the observed ratio of \ion{C}{3}] to 
\ion{N}{3}] brightness is consistent with an abundance ratio  
$n_C/n_N \approx 0.1$.
The width of \ion{C}{3}] $\lambda$1909 decreased by a significant 
amount; its FWHM was roughly 240 km s$^{-1}$ in 2002-2004 but only 
160 km s$^{-1}$ in 2013-2015.   

    \vspace{1mm}  

\ion{Si}{3}] $\lambda$1892 should originate in ionization zones 
of H$^+$ and He$^0$.  It is the second most conspicuous 
emission line between 1300 and 2400 {\AA} 
(Fig.\/ \ref{fig:uvtracing}), and appears to have at least two 
components in the range $-300$ to $-120$ km s$^{-1}$.  Since it can be 
formed via an exotic two-photon process \citep{jo06}, this line may 
convey unique information about the radiation field.  

    \vspace{1mm}  

These three semi-forbidden features -- \ion{N}{3}], \ion{C}{3}], and 
\ion{Si}{3}] -- share a suggestive velocity trend.   
Vertical dashed lines in Figure \ref{fig:1900A}   
assist in seeing this effect.  In 2002 and 2004, the left sides 
of \ion{Si}{3}] and \ion{C}{3}] in the figure had wavelengths 
about 0.6 {\AA} smaller than in 2013 and 2015 -- i.e., at the earlier 
times, some material was approaching us about 100 km s$^{-1}$ faster.   
A careful comparison of Figures 
\ref{fig:niii2002} and \ref{fig:niii2013} shows the 
same effect for the main peak in the \ion{N}{3}] multiplet.   This 
velocity change is substantial, even though it appears inconspicuous 
in the figures.  If these were permitted lines, P Cyg absorption 
would obscure that part of the emission profile. 

   \vspace{1mm} 

The above effect might indicate a diminished wind speed;   
but another, more interesting explanation involves the geometry 
of the shocks.   Consider an emission line that originates 
in or near the shocked region, in gas flowing 
roughly parallel to a pseudo-hyperboloid surface like 
boundary \underline{3-4} in Figure \ref{fig:zones}. 
(In {\S}\ref{subsec:niiid} we called it an umbrella.)      
Further suppose that our line of sight direction (L.O.S.) was 
just within the opening angle of the pseudo-hyperboloid at those 
times -- i.e., HST viewed the umbrella obliquely from its 
concave side.  Then the most extreme negative Doppler 
velocities originate in the region where the flow is almost 
anti-parallel to the L.O.S.\/ -- the near side of the umbrella.        
Due to gradual weakening of the primary wind, the shock's opening 
angle should have widened between 2002 and 2015.  
This trend increased the projection angle between the 
flow velocity and the anti-L.O.S.;  thereby causing the observed 
effect on the short-wavelength side of the Doppler profile. 
An increase of the opening angle can have a similar effect on 
emission from primary wind region \underline{3} in Figure \ref{fig:zones}, 
for partially different reasons.  If an explanation     
in this vein is correct, then the total opening angle of the 
shocked zone must have increased by at least 20$^\circ$.  

    \vspace{1mm}  

Alternative explanations can be devised, but in any case 
this velocity effect is one example of the range of questions 
that a realistic {\it evolving } model needs to answer. 

  \vspace{1mm}  


    \vspace{3mm}

\section{Absorption lines and vanishing material}      
\label{sec:abslines}  

   \vspace{1mm}  

Unlike visual wavelengths, the UV has many permitted transitions from 
well-populated levels near the ground states.  Even small amounts of  
material can thus form strong absorption features.  In $\eta$ Car's 
spectrum, they show rapidly declining column densities of 
several ion species.  

     \vspace{1mm}  

Low-ionization absorption lines can occur at large distances from the 
primary star,  and many component velocities have been listed, e.g., 
by \citet{gu06}.  In the time 
interval from 2002 to 2015, material flowing outward from $\eta$ Car 
could move through distances of the order of 1000 AU.   
Here we examine only a few definite features 
seen at $t = 2002.51$ and 2013.82.  Other lines, and the data from 2004 
and 2015, are  consistent with these findings.

 \begin{figure}   
 \plotone{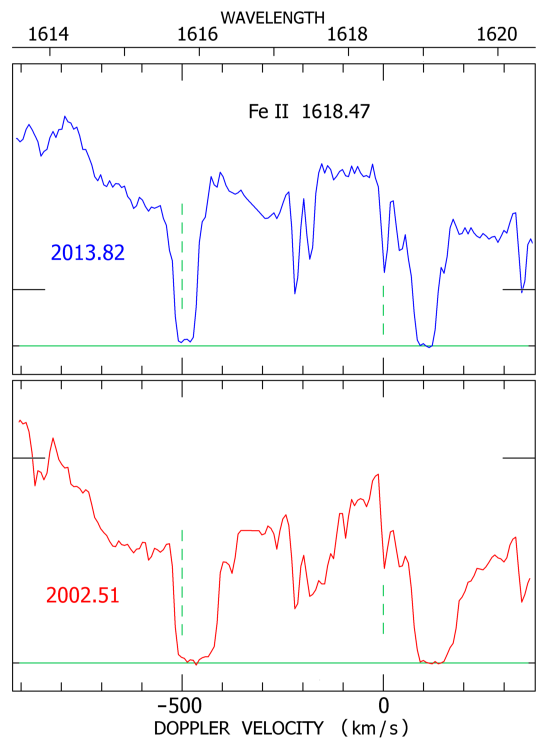} 
 \caption{The \ion{Fe}{2} $\lambda$1618 absorption feature 
    in 2002 and 2013.  The zero-velocity absorption is interstellar, 
    and the broad absorption on the right side is a different line.  	
    Horizontal marks at the sides indicate 
    $f_\lambda = 4 \times 10^{-12}$ erg cm$^{-2}$ s$^{-1}$ {\AA}$^{-1}$,   
    not corrected for extinction. }  
 \label{fig:1617A}    
 \end{figure}

   \vspace{1mm}  

Figure \ref{fig:1617A} shows \ion{Fe}{2} $\lambda$1618 absorption. 
One vertical line marks the lab wavelength and interstellar absorption, 
while another indicates Doppler velocity $-500$ km s$^{-1}$.  
Very strong  absorption extended from $-520$ to $-420$ km s$^{-1}$ in 
2002, but eleven years later the longer-wavelength half of this 
interval had become nearly transparent.  Column densities also 
declined in velocity ranges $-420$ to $-360$ km s$^{-1}$ 
and $-200$ to $-100$ km s$^{-1}$.  Quantitative assessments would     
be complicated, but Figure \ref{fig:1617A} strongly suggests that  
{\it the optical depth fell to less than half of its initial value\/}   
at most velocities between $-470$ and $-100$ km s$^{-1}$.  A similar 
decrease may have occurred between $-520$ and $-470$ km s$^{-1}$, 
but if so it was undetectable because the optical depth there 
remained large.  

   \vspace{1mm} 

\ion{Ni}{2} $\lambda$1752, which obscures the \ion{N}{3}] profile 
at 1749 {\AA}, exhibits the same tendencies;  compare Figures 
\ref{fig:niii2002} and \ref{fig:niii2013}.  This weaker line 
shows that the amount of material at $-350$ to $-100$ km s$^{-1}$ 
was small compared to the main outflow;  but here we emphasize  
the consistent trends.  Near the left edge of 
Figures \ref{fig:niii2002} and \ref{fig:niii2013} one can see a 
stronger \ion{Ni}{2} line with similar changes.  These effects 
appear to be consistent among the numerous low-ionization UV 
absorption lines.  

    \vspace{1mm} 

The strong \ion{Al}{2} $\lambda$1671 and \ion{Al}{3} 
${\lambda}{\lambda}$1855,1863 lines have unique qualities for 
this subtopic.  Al$^+$ coexists with H$^0$ and Fe$^+$, while Al$^{++}$ 
coexists with H$^+$ and Fe$^{++}$;  but these aluminum ions have much 
simpler spectra than \ion{Fe}{2} and \ion{Fe}{3}.  Instead of hundreds 
of confused features, they show only the three lines named above -- 
all arising from ground level, with strong oscillator strengths 
$f_{ij} \approx$ 1.77, 0.56, and 0.28.  Moreover, we can 
employ the fact that \ion{Al}{3} $\lambda$1855 is 
intrinsically twice as strong as $\lambda$1863.  

    \vspace{1mm}  
 \begin{figure}   
 \plotone{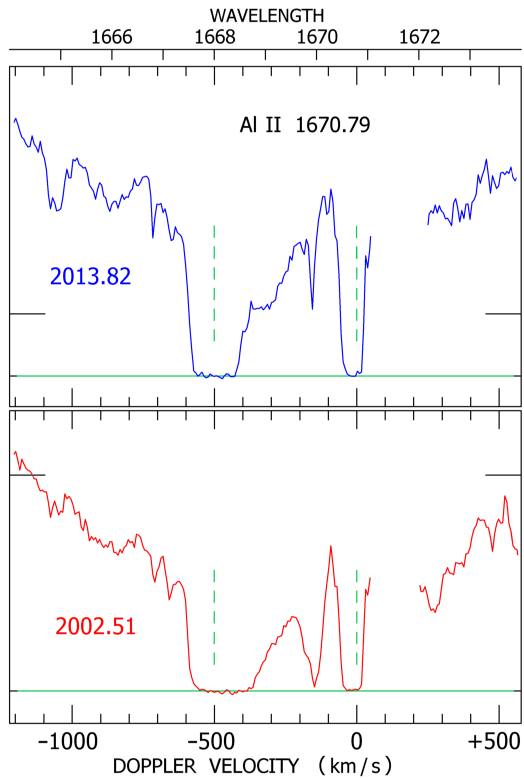} 
 \caption{The \ion{Al}{2} $\lambda$1671 absorption feature in the years 
    2002 and 2013.  Horizontal marks at the sides indicate 
    $f_\lambda = 4 \times 10^{-12}$ erg cm$^{-2}$ s$^{-1}$ {\AA}$^{-1}$,   
    not corrected for extinction.  The data have a gap around 
    1671.5 {\AA} because two echelle orders did not overlap. } 
 \label{fig:1670A}    
 \end{figure}
 \begin{figure}   
 \plotone{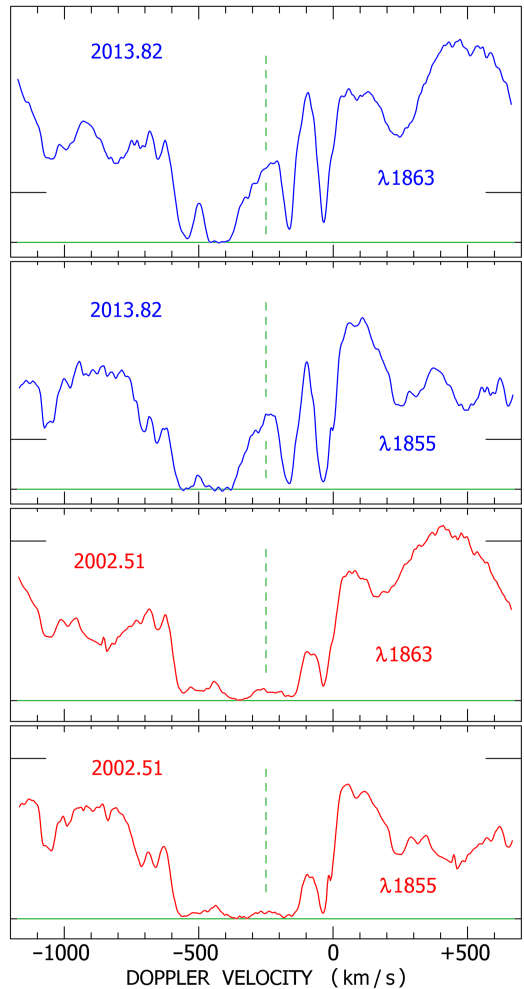}  
 \caption{\ion{Al}{3} ${\lambda}{\lambda}$1855,1863 absorption  
     features seen in 2002 and 2013.  The vertical dashed line at 
     $-250$ km s$^{-1}$ is a guide for noting a subfeature 
     noted in the text.  Horizontal markers at the sides indicate 
     $f_\lambda = 4 \times 10^{-12}$ erg cm$^{-2}$ s$^{-1}$ {\AA}$^{-1}$, 
     not corrected for extinction. }     
 \label{fig:aliii}    
 \end{figure}

These features are shown in Figures \ref{fig:1670A} and 
\ref{fig:aliii}, and together they support a particular 
chain of reasoning.  First, the changes in \ion{Al}{2} 
$\lambda$1671 were consistent with the \ion{Fe}{2} and \ion{Ni}{2} 
trends;  absorption decreased at velocities between 
$-420$ and $-150$ km s$^{-1}$.  But the next point is less 
routine:  {\it The \ion{Al}{3} profile in 2013 nearly matched 
that of \ion{Al}{2} in 2002!}  (Compare the two figures. 
The $\lambda$1671 tracing has a data gap between $+50$ and 
$+200$ km s$^{-1}$ which probably hides an emission peak, 
but this does not affect its absorption part.)  If we view only 
this striking resemblance by itself, an obvious interpretation 
is that  some gas became more highly ionized.  

    \vspace{1mm}

However, the \ion{Al}{3} lines then lead to an intricate 
puzzle with potentially major implications.  Figure \ref{fig:aliii} 
shows a declining, not increasing, column density of 
Al$^{++}$.  In both 2002 and 2013, that quantity appears to have 
been {\it larger} than the Al$^+$ column density in the velocity range 
where a difference can be detected.  Now, consider the local brightness 
maximum (i.e., absorption minimum) between $-350$ and 
$-200$ km s$^{-1}$, marked by vertical lines in Fig. \ref{fig:aliii}.  
It was conspicuous in \ion{Al}{2} in 2002, and in \ion{Al}{3} in 2013.   
Judging from the overall appearance 
of this subfeature,  one would naively guess that optical depths 
in that velocity range were moderately small when it was visible. 
However: both \ion{Al}{3} lines represent precisely the same 
material, and the $\lambda$1855 line has an oscillator strength 
twice as large as for $\lambda$1863. Therefore the optical depth 
$\tau(-250 \; \mathrm{km} \; \mathrm{s}^{-1})$ should have    
been visibly larger in the $\lambda$1855 feature. Hence 
the bump should have been appreciably fainter or less prominent 
in the \ion{Al}{3} $\lambda$1855 line compared to $\lambda$1863, 
contrary to  observations.  

   \vspace{1mm} 

How can we explain this discrepancy?  A hypothetical ``extra emission'' 
source for the bump would be implausible, since the emissivities differ 
greatly among the three \ion{Al}{2} and \ion{Al}{3} lines.  
The UV halo \citep{hi06} does not provide a simple explanation, since 
the two \ion{Al}{3} features almost certainly depend on optical depth 
$\tau(v)$ even if it refers to scattering paths 
rather than a direct line of sight.  If this were not true, then the 
changing structure in Figure \ref{fig:aliii} would be very hard to   
understand.  Hence the $\lambda$1863 line should appear different from 
$\lambda$1855 even if one cannot predict the details.\footnote{ 
  The arguments concerning optical depth in {\S}\ref{subsec:contin_a} do 
  not apply here, because they applied only to the general appearance 
  of the \ion{Fe}{2} forest with fairly large optical depths.  The 
  problems in this section refer, instead, to particular velocity 
  intervals in particular lines, where the optical depths ceased 
  to be large. } 
We propose the following idea, which significantly affects the  
small-scale geometry of the situation.
Suppose that the brightness in these absorption lines between $-100$ and 
$-350$ km s$^{-1}$   depends chiefly on area covering factors, rather than 
a simple optical depth.  In this scenario, part of the projected area 
is quite dark at those wavelengths because it is optically thick, 
but another part is almost free of Al$^+$ and Al$^{++}$.    
Since velocities depend on location, the area covering factor is 
a function of Doppler velocity. 
The resulting $\lambda$1855 and $\lambda$1863 absorption 
profiles should look alike, as observed.  With a few 
additional assumptions, this concept may partially explain 
the resemblance   between \ion{Al}{2} in 2002 and \ion{Al}{3} 
in 2013.  Note that this problem involves tiny amounts of material 
in the stated velocity range.  Since the required optical depths 
imply scarcely more than 0.1\% of the expected outflow of 
aluminum ions in our direction, perhaps we should be 
surprised that these absorption lines are not dark at all 
wavelengths from zero to $-600$ km s$^{-1}$. 

     \vspace{1mm} 

In summary, the above discussion has an obvious underlying thread:
{\it In each absorption feature, at almost every velocity where 
the optical depths are suitable for detecting a change in column 
density, a decrease was seen} -- and it generally appeared to be 
of the order of 50\% in 11 years, not 10\% or 20\%. 
This was not merely an ionization effect, since it includes both  
Al$^+$ and Al$^{++}$.  We can safely assume that these represent 
most of the aluminum in the absorption-line regions, because 
the next ionization stage Al$^{+3}$ would require a considerable 
flux of helium-ionizing photons above 28 eV, at distances far 
from the star.       

    \vspace{3mm}

\section{Summary} 
\label{sec:sum}  

Above we have presented an initial survey of some particular 
HST UV observations of $\eta$ Car.  The results fall into 
three categories;   
(1) the star's rapid change of state, 
(2) morphology of gas flows across a wide range  of size scales, and 
(3) strange features that are not easy to explain.  

\subsection{$\eta$ Car's rapid secular trends}  
\label{subsec:sum1}  

As noted in {\S}\ref{sec:intro}, we consider the change of state 
to be the main development in this topic since 2000, because it 
relates to   stellar structure and giant-eruption instability.   The 
UV results outlined in this paper expand the variety of evidence.      

    \vspace{1mm} 

Figures 4-6, 10, and 12-14 show many large changes in 2002-2015.  
Circumstellar UV extinction decreased by 30\% or more along our
line of sight
({\S}\ref{sec:contin} above).  Absorption lines indicate that column 
densities greatly decreased at almost every Doppler velocity where 
a change would be detectable, and practically vanished at some 
velocities ({\S}\ref{sec:abslines}).  This statement includes Al$^{++}$, 
not just singly-ionized species.  These results are consistent with 
trends seen at longer wavelengths \citep{mar06b,mar10,me10b,me12,da15}.    
In addition, the short-wavelength sides of some emission line profiles 
shifted between 2004 and 2013 ({\S}\ref{sec:emlines}).  

    \vspace{1mm} 

If one wishes to explain the observed facts without a major change in 
$\eta$ Car's wind, it is necessary to invoke multiple hypotheses. 
$(a)$ Our line of sight must be abnormal in showing major decreases 
in extinction and line absorption.  
$(b)$ Low-excitation emission lines -- whose appearance depends 
only mildly on viewing direction --  must be very sensitive to minor 
changes in the star and its wind.  
$(c)$ Some third idea must account for spectacular changes in the 
nature of periastron events from 1998 to 2014 \citep{da05,da15,me11,me15}.  
$(d)$ Velocity profiles of semi-forbidden emission lines (which are not 
complicated by P Cyg absorption) changed in a particular way noted in 
{\S}\ref{sec:emlines} above.   
Each of these points can individually be explained with its own  
special hypothesis, but we would need to accept all of them  
together.  

   \vspace{1mm}

{\it It is simpler to deduce, instead, that the stellar wind has
diminished by a significant amount} -- e.g., by roughly 50\% 
since the earliest STIS observations in 1998.  That one 
hypothesis, perhaps including a change in the latitude dependence,
can explain nearly all of the observed effects -- see 
\citet{da12,me12,mar06a} and references therein.  It 
strongly suggests a progressive alteration in the stellar 
radius and/or surface rotation and/or luminosity.  

     \vspace{10mm}     

\subsection{Morphology of the emission regions, and numerical simulations}  
\label{subsec:sum2} 

Eta Car produces an extraordinary amount of \ion{N}{3}] $\lambda$1750 
emission, carrying about as much energy as the entire stellar wind. 
In {\S}\ref{subsec:niiib} we used the \ion{N}{3}] multiplet ratios
to estimate a 
characteristic gas density $n_e \sim 10^{11}$ cm$^{-3}$,  about two 
orders of magnitude denser than the primary wind in that vicinity.  
The high density implies that \ion{N}{3}] $\lambda$1750 
originates in condensations within the colliding-wind shocked region.  
If our density estimate is seriously wrong, then the \ion{N}{3}] 
and \ion{He}{1} emission lines probably occur in a region of the 
primary wind, zone  \underline{3} in Figure \ref{fig:zones}.      

   \vspace{1mm} 

As outlined in {\S}\ref{subsec:niiic}, thermal instability should 
produce small dense ``cloudlets'' in the shocked gas.  The observed 
strength of \ion{N}{3}] emission, combined with the high density noted 
above, appears to confirm that this does happen and shows that the 
cloudlets are not immediately destroyed.  Roughly $10^5$ to 
$10^7$ cloudlets are required in order to convert most of the secondary 
star's helium-ionizing photons into \ion{N}{3}] and other emission. 
(Filamentary condensations can be regarded as strings of cloudlets.) 

    \vspace{1mm} 

Thus, a valid global model of the shocked region requires more 
than $10^8$ adaptive sample points,  with many time steps  
in the densest locales.  A more practical approach, of course, is 
to produce local simulations of a few hundred cloudlets and combine 
them with a lower-resolution global flow model -- but we are not 
aware of efforts like that for this object.  Published accounts of 
calculations for $\eta$ Car's shocked region (e.g. \citealt{cl15}) 
have not described the small-scale morphology with densities  
$n_e \sim 10^{11}$ cm$^{-3}$.  

     \vspace{1mm} 

Realistic models will need additional effects.  For instance, both winds 
are generally thought to be inhomogeneous with dense ``clumps.''  When 
a clump enters the shocked region, it may penetrate to a considerable 
distance like a raindrop falling on soft snow, with a density-enhanced 
cooling rate.  On the secondary wind side, this effect should    
increase the observable X-ray luminosity and tends to 
destabilize the shock front.  On the primary side, it may determine 
the size, number, and morphology of cloudlets.  A simple filling 
factor is not adequate for modeling the winds!  Meanwhile, since 
cloudlet formation entails a large density compression factor, 
magnetic pressures and tensions are likely to be significant.  
Models of the primary wind are presumably more robust, but similar 
doubts apply to them as well. They neglect details of the inhomogeneities 
and other phenomena, and they have simplified geometries.   
In summary, {\it existing numerical simulations of $\eta$ Car's outflows 
should be regarded as preliminary sketches, not accurate models.}

    \vspace{1mm}  

In {\S}\ref{subsec:niiid} and {\S}\ref{sec:emlines} we mentioned 
some details that relate to large-scale morphology.  They concern 
emission line profiles and very likely the opening angle 
of the shocked region;  see those sections.  Velocity components  
within the absorption lines ({\S}\ref{sec:abslines}) represent 
structures in the outflow, but without other information their 
locations are conjectural.

\subsection{A major unexplained feature}  
\label{subsec:sum3}

In 2003-2004, there was a huge amount of  \ion{N}{3}] emission 
at velocities around $+800$ km s$^{-1}$ 
({\S}\ref{subsec:niiie} and Fig. \ref{fig:niii2002}).  Such a large 
{\it recession} velocity is unusual in $\eta$ Car, and the  
associated luminosity was extraordinary.  We suspect that it 
was related to a burst of mass ejection in the 2003.5 periastron 
event (cf.\ \citealt{mar06a}), but we do not understand its 
excitation.  A similar but weaker feature probably existed 
in 2015 (Fig. \ref{fig:niii2013}),  and possibly in 2002.   

    \vspace{1mm}

Helium-ionizing EUV from the hot secondary star is probably the 
only available energy source for this type of emission, but it 
requires a clear path between that star and the receding gas on 
the far side of the primary, with no intervening part of the 
primary wind.  The system's large-scale spiral pattern might 
provide such a path 
-- see, e.g., the fourth panel in Figure 3 of \citet{pa09}.  
However, only a very small fraction of the secondary star's EUV  
goes in that direction, so the energy budget is doubtful.    

    \vspace{1mm}

Theorists who enjoy unorthodox models may be able to devise  
other ways to excite this strange  \ion{N}{3}] emission. 
For instance, the latitude-dependent primary star might 
conceivably have a hot equatorial zone providing EUV. 
Our main point here, though, is twofold:  This feature   
constitutes a fascinating puzzle, and its high velocities   
may be significant regarding the periastron 
spectroscopic event. 

    \vspace{1mm}

Finally, we emphasize two facts. (1) This paper has been essentially 
a reconnaissance.  Many features in the data have been omitted here, 
and no elaborate models have been employed.      
(2) HST observed $\eta$ Car in the UV on several other occasions, 
ignored here because their orbit phases were unsuitable for 
our present purpose.   {\it The entire HST UV  data set 
on this object contains enough material for many investigations  
in a variety of interesting and significant problems.}  

     \vspace{4mm}



   ---   ---   ---   
   \vspace{1mm}

{\it Acknowledgements --}  
We are grateful to A. Mehner for many contributions;  she was the P.I.  
for HST programs GO 13377 and 13789 which obtained the UV spectra in  
2013 and 2015.  As always, we appreciate the excellent help given by  
B. Periello and other STScI staff members in planning and scheduling 
the HST observations.  K.I. was supported by Grant-in-Aid for 
Scientific Research (C) (JSPS KAKENHI Grant Number JP26400227).

\end{document}